\documentstyle[prl,aps,psfig,twocolumn]{revtex} % COMMENT OUT IF
\begin{document}
\draft
%                            % COMMENT OUT NEXT LINE IF PREPRINT
\twocolumn[\hsize\textwidth\columnwidth\hsize\csname
@twocolumnfalse\endcsname
\title{Universal $1/f$ Noise from Dissipative SOC Models.}
\author{Paolo De Los Rios and Yi-Cheng Zhang}
\address{Institut de Physique Th\'eorique, 
Universit\'e de Fribourg, CH-1700, Fribourg, Switzerland.}
\date{\today}
\maketitle

\begin{abstract}

We introduce a model able to reproduce the main features of $1/f$ noise:
hyper-universality (the power-law exponents are independent on the dimension
of the system; we show here results in $d=1,2$) 
and apparent lack of a low-frequency cutoff in the power 
spectrum. Essential ingredients of this model are an activation-deactivation
process and dissipation. 

\end{abstract}
]
\narrowtext

The voltage drop $V$ on a resistor of resistance $R$ through which 
a current $I$ is flowing
obeys the well known ohmic law $V = R I$. 
Yet, when we look carefully,
we discover that such a voltage is not perfectly constant through time.
Indeed there are noise fluctuations around $V$.
The spectral density of these fluctuations clearly shows 
a $1/f$ behavior on many decades in the frequency domain.
This is a well known example of $1/f$ noise, one of the most
common and widespread features in nature.
It appears in a variety of systems ranging from the light of quasars
\cite{Press78} to water 
flows in rivers\cite{MW68}, music and speech\cite{VC75}, 
and the already mentioned electrical 
measurements\cite{DH81,Weissman88}.
Despite its ubiquity and universality, a clear and simple explanation for
such a behavior is still lacking. Indeed, it is possible to find in
the literature some {\it ad hoc} formulas and theories, but most of them
are based on unverified assumptions, or they catch a glimpse of the physics
only of some particular system, therefore missing to
address the widespread occurrence of the phenomenon\cite{Weissman88}.

In the search for a universal mechanism of $1/f$ noise, 
Bak, Tang and Wiesenfeld (BTW) proposed the new concept of self-organized 
critical (SOC) systems\cite{BTW87}: These are systems driven by their own dynamics
to a state characterized by power-law time and space correlations, and 
therefore also by power-law ($1/f^\alpha$) power spectra.
Yet, a number of features of SOC systems do not show agreement with
the features of $1/f$ noise: the exponent $\alpha$ is seldom close to $1$, 
and it depends strongly on the dimensionality of the system (at least below
the upper critical dimension, which is in general high\cite{LU97,CMZ98,DMV98}); 
moreover, 
in SOC systems power-law time correlations are always found in the presence of
power-law (long range) space correlations, for which there is no evidence
in most systems exhibiting $1/f$ noise\cite{nocorrelations}.

In this Letter, we propose a simple model, inspired by a SOC model originally
introduced by one of us\cite{Zhang89}, able to implement some of the current and most 
accepted ideas on $1/f$ noise, and to show a clear $1/f$ behavior
independent on the dimension of the system (therefore ``hyper-universal''). 

The basic model is a continuous version of the BTW sandpile. Given a lattice,
to every site $i$ is associated a continuous variable $x_i$
(representing, say, energy). The basic 
time step of the dynamics consists in changing 
the value of an energy $x_i$ of a positive random 
quantity $\epsilon$ (taken from some probability distribution $P(\epsilon)$)
\begin{equation}
x_i(t+1) = x_i(t) + \epsilon(t)
\label{adding}
\end{equation}
with the $\epsilon(t)$ variables uncorrelated in time.
Whenever this addition step makes an energy 
$x_i$ greater than a certain value $x_c$, then 
the quantity $x_i$ is redistributed to the $2 d$ 
nearest neighbors of site $i$
\begin{equation}
x_j(t,\tau+1) = x_j(t,\tau) + \frac{x_i(t,\tau)}{2d}
\label{redistribution}
\end{equation}
and the energy $x_i$ is reset to $0$.
The time variable $\tau$ is used to describe the redistribution
process, which is considered to be much faster than 
the process of addition (\ref{adding}). 
It is possible that this redistribution drives some other
energies to exceed $x_c$, triggering new redistributions.
This process (an {\it avalanche} in the jargon) 
goes on as long as there are no more
energies greater than $x_c$. Then a new quantity is added,
as in (\ref{adding}), and time $t$ in increased by $1$.
It is important to remind the presence of two different time-scales:
a slow one, corresponding to the addition of energy,
and a fast one, corresponding to the redistribution of energies
which are above $x_c$.
The statistical properties of this model after a transient time
are very interesting.
The distribution of the energies on the lattice clearly shows
a {\it quantization}, with peaks around $0$, $x_c/2d$, $2x_c/2d$,...,
$(2d-1) x_c/2d$. Moreover the distribution of avalanches
with respect to their duration (measured on the internal time variable $\tau$)
obeys a power-law. In general it is possible to show
that there are long-range (power-law) space and time correlations.
Indeed, this is a SOC model. A further quantity that is interesting to 
look at is the total lattice energy content  $X(t) = \sum_i x_i(t)$. 
It represents a signal whose power spectrum also obeys some
power-law in the frequency domain. 

This model (as well as the original BTW sandpile) is mainly an 
activation/deactivation process,
which is believed to be one of the
main features relevant for the
description of $1/f$ noise\cite{MMM93}.
Yet, as pointed out above,
the exponents depend on the dimensionality of the lattice, and are never
close to $1$\cite{Zhang89}. Therefore the model, as it stands, is not
a good candidate to describe $1/f$ noise. 

New ingredients need to be added to the model: the energy is added only on 
one side of the lattice, defining implicitly a preferred 
propagation direction for the energy; the second ingredient is dissipation.
During redistribution, 
we added some dissipation in the form
\begin{equation}
x_j(t,\tau+1) = x_j(t,\tau) + \frac{x_i(t,\tau)}{2d}(1-a)
\label{dissipation}
\end{equation}
With this new rule avalanches cannot establish anymore long-range correlations
throughout the system and are not anymore power-law distributed. In a word, 
dissipation destroys the self-organized criticality of the system.
Yet, some features, such as the {\it quantization} of the energy levels, 
survive.

\begin{figure}
\centerline{\psfig{file=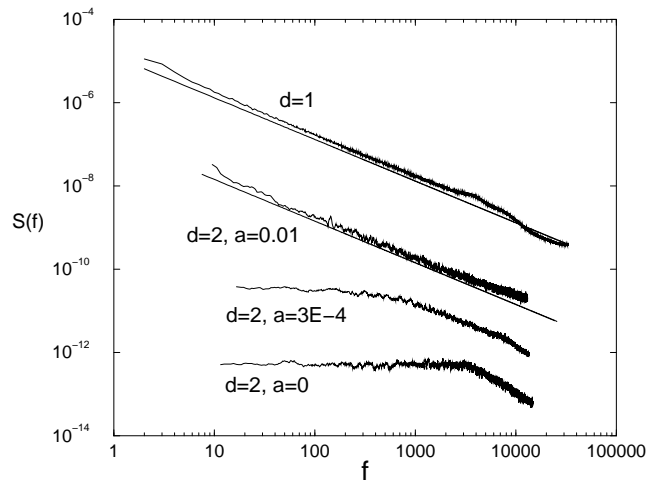,width=8.5cm,angle=0}}
\caption{Log-log plot of the power spectrum of 
$X(t)$ in $d=1$ and $d=2$. The straight
lines are $1/f$ power-laws drawn for reference.
The system size used for the $d=1$ simulations is $L=100$, with
$\epsilon_{max}= 10$ and $a=0.03$; $d=2$ simulations are performed with
$L=100$, $\epsilon_{max} = 1$ (energy is added on all the sites
$x_{0,i}$, for a maximum possible energy injection of $100$) and $a=0.01$. 
For comparison, we also add the $d=2$ 
power-spectrum 
with a much smaller dissipation ($a=0.0003$, 
but with the same 
size and energy injection regime as before)
that
shows crossover between clean $1/f$ behavior,
and the behavior in the absence of dissipation (lowest curve)
that clearly shows no sign of $1/f$ behavior.}
\label{Fig: fig1}
\end{figure}

In our implementation of the model, we inject energy on one side of 
the lattice according to (\ref{adding}), and let it propagate through the
lattice following (\ref{dissipation}). We compute the power spectrum of 
$X(t)$, finding a clear $1/f$ behavior both in $1$ and $2d$ (see
Fig.\ref{Fig: fig1}) for at least three decades. This is a signature of the 
desired (and observed in nature) ``hyper-universality''.

Of course the details of the implementation are relevant up to
some level: We take the added random energy $\epsilon$
from a uniform distribution in $[0,\epsilon_{max}]$, with
$\epsilon_{max} \gg x_c$. Indeed, if $\epsilon_{max} \le x_c$
then, on the average, it takes some steps before the energy goes
above threshold. Since $\langle \epsilon(t) \epsilon(t') \rangle
= \delta(t-t')$, this implies that short time fluctuations of $X(t)$
are uncorrelated and the resulting power spectrum has a flat tail
for high frequencies. Moreover, we find that for larger systems the
frequency range where $1/f$ behavior emerges is broader; yet,
due to the non criticality of the model, the transient time
to go to stationarity grows fast with the system sizes,
forbidding us to explore systems larger than $128 \times 128$ lattice sites
in two dimensions. Also, dissipation cannot be too small: indeed,
when $a$ becomes very small, then we approach the SOC system, which is
characterized by different exponents, and crossover effects emerge.
In Fig.\ref{Fig: fig1} we show the $d=2$ power-spectrum with
different dissipation regimes. In the absence
of dissipation, the power-spectrum is essentially flat, 
with only a small region of 
power-law behavior $1/f^{1.5}$. A very small dissipation ($a=0.0003$)
clearly gives an intermediate behavior, with a high frequency 
$1/f$ region, and a low frequency flat one.
Of course, the larger the dissipation, the larger the energy 
injection must be in order to acivate all the lattice.

We believe nonetheless that the way crossover emerges is model-dependent.
Indeed we believe that the relevant features of the model are
nonlinearity (in this case, activation/deactivation of sites)
and dissipation.

Additionally, we investigate the power spectrum $S(f,x)$ of the
energy at site $x$. We find that, for large $x$, $S(f,x)$ has a scaling form 
\begin{equation}
S(f,x) = e^{\delta x} h(f e^{\delta x})
\label{scaling form}
\end{equation}

From this scaling form we can infer that there is a characteristic 
time $T(x) \sim e^{\delta x}$ 
associated with a site at distance $x$ from the origin of
the lattice; we can build an intuitive picture of this characteristic time
thinking that in order for the energy to propagate
from site $x$ to site $x+1$, it has to overcome some barrier,
with a characteristic time to overcome it taken as an Arrhenius law
$e^{\delta}$. Then, in order to propagate from the origin down
to site $x$, the characteristic time becomes, roughly, of the
order of $e^{\delta x}$.  
As an alternative explanation,  due to dissipation, 
energy has a probability to 
propagate to a depth $x$ which is exponentially decreasing with $x$,
hence an exponential characteristic time associated with $x$.

Due to dissipation, there are no long range correlations in the system.
Therefore, as a first approximation, the energies in different
sites are uncorrelated. The total power spectrum can 
therefore be written as
\begin{eqnarray}
S(f) & = & \sum_x S(f,x) \sim \int_0^L e^{\delta x} h(f e^{\delta x}) dx 
\nonumber \\
& = & \frac{1}{\delta f} \int_0^{f e^{\delta L}} dy h(y)
\label{spettro 1/f}
\end{eqnarray}
(indeed the power spectrum of uncorrelated signals is just the sum
of the power spectra of the signals, a signature of linear 
superposition).

\begin{figure}
\centerline{\psfig{file=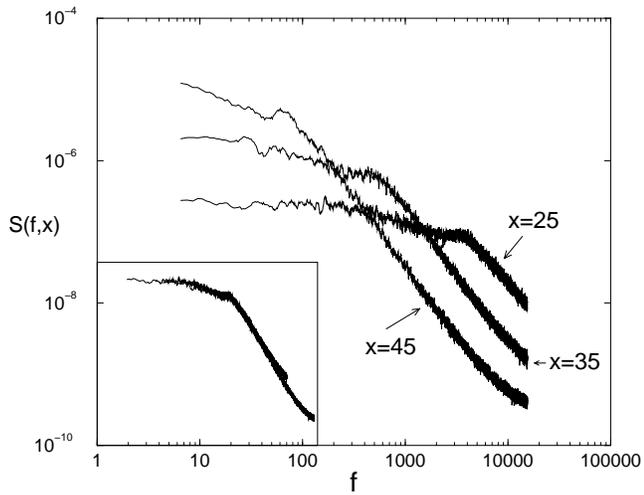,width=8.5cm,angle=0}}
\caption{Log-log plot of the power spectra $S(f,x)$ for $x=25,35,45$ on a 
$d=2$ lattice of $L=100$, $\epsilon_{max} = 10$ and $a=0.01$. 
The large $f$ behavior has a characteristic slope
$1/f^2$. The collapse in the inset is obtained plotting $e^{-\delta x} S(f,x)$
vs. $f e^{\delta x}$ with $\delta = 1.9$. Indeed the universality of the
scaling function ($4$) emerges. 
}
\label{Fig: fig2}
\end{figure}

We see therefore that the $1/f$ behavior of the power spectrum emerges
as the superposition of local power spectra that have nothing
to do with $1/f$ noise. The lower cutoff frequency $f_c \sim e^{-\delta L}$
vanishes extremely fast in the thermodynamic limit, accounting for the observed
experimental absence of a lower cutoff (whose presence is necessary
to have a finite power associated with the signal).
Such a superposition mechanism to obtain a $1/f$ power spectrum
is strongly reminiscent of the McWhorter model\cite{Weissman88,McWhorter57}: $S(f) =  \int S(f,f_c) P(f_c) df_c$
with $P(f_c) \sim 1/f_c$ the distribution of the frequencies $f_c$ and
$S(f,f_c) \sim f_c/(f^2+f_c^2)$. In our model we have $df_c /f_c = dx$,
accounting for the correct distribution of characteristic 
frequencies, but we have no explicit form for $S(f,f_c)$ (although also
in our case the large $f$ behavior is $1/f^2$, see Fig.\ref{Fig: fig2}).   
The relation $f_c = T_c^{-1} = e^{-\delta x}$ can have many different
underlying physical origins, as diverse as tunnelling between different traps,
jumps between metastable states distributed in space or dissipation
in an activation/deactivation process
(as in the present realization). Such a variety of mechanisms giving rise
to the good scaling functions (and many other may be concieved) 
strongly points to the observed widespread occurrence of $1/f$ noise
in nature.

Actually, dissipation associated with SOC was already considered in
\cite{MKK90,COB92} and more recently in \cite{VZ97}. 
There
non-trivial power-laws in the power spectrum where found, strongly dependent
on the dissipation coefficient. Moreover dissipation and driving
where chosen in such a way that the SOC behavior was not destroyed.
Actually, many experiments
show that there is indeed universality, that the power spectrum
is close to $1/f$, and that the long range time and space correlations
typical of SOC systems are absent\cite{nocorrelations}. 
All these ingredients, on the contrary,
are present in our model, where changing the parameter values gives rise
to crossover effects but not to non-universality.

Our model shows moreover the relevance of a propagation direction
(feeding energy from one side of the system, and extracting it from the 
other), and the $1/f$ power spectrum turns out to be the
superposition of power spectra that are far from $1/f$.

In conclusion, we have introduced dissipation in a well-known 
SOC model: As a consequence, the critical behavior of the system is destroyed
(but not its self-organization properties, such as the quantization of the
energy levels). The resulting power-spectrum has a clean $1/f$ behavior both
in $1$ and $2$ spatial dimensions: We believe that our $1$ and $2$
dimensional results
hint toward the independence
of such behavior on the actual dimensionality of the system, 
and to the desired hyper-universality.
We also unveiled that the origin of such a behavior has to be found
in the superposition of power-spectra with characteristic frequencies
$f_c$ suitably distributed in space. Such a distribution is not
an input in the model, but emerges due to the directedness
properties of our model. The desired distribution
of characteristic times typical of the McWhorter
model emerges spontaneously in our model.
One can concieve many other situations where
the distribution of the characteristic frequencies is instead 
given, and the system only has to cope with it.
As a consequence, the present model not only provides a simple 
and hyper-universal explanation of $1/f$ noise, but is also
suggestive of a wide variety of microscopic physical
mechanisms able to give $1/f$ noise.

This work has been partially supported by the European network
contract FMRXCT980183.

\begin{thebibliography}{99}

\bibitem{Press78} W.H. Press, Comments Astrophys. {\bf 7}, 103 (1978).

\bibitem{MW68} B.B. Mandelbrot and J.R. Wallis, Water Resour. Res. {\bf 4},
909 (1968); {\bf 5}, 321 (1969).

\bibitem{VC75} R.F. Voss and J. Clarke, Nature {\bf 258}, 317 (1975);
J. Acoust. Soc. Am. {\bf 63}, 258 (1978).

\bibitem{DH81} P. Dutta and P.M. Horn, Rev. Mod. Phys. {\bf 53}, 497 (1981).

\bibitem{Weissman88} M.B. Weissman, Rev. Mod. Phys. {\bf 60}, 537 (1988).
 
\bibitem{BTW87} P. Bak, C. Tang and K. Wiesenfeld, Phys. Rev. Lett. {\bf 59},
381 (1987); Phys. Rev. A {\bf 38}, 364 (1988).

\bibitem{LU97} S. Lubeck and K.D. Usadel, Phys. Rev. E {\bf 56}, 5138 (1997).

\bibitem{CMZ98} A. Chessa, E. Marinari and S. Zapperi, Phys. Rev. E 
{\bf 57}, R6241 (1998).

\bibitem{DMV98} P. De Los Rios, M. Marsili and M. Vendruscolo, Phys. Rev. Lett.
{\bf 80}, 5746 (1998).

\bibitem{nocorrelations} R.D. Black, M.B. Weissman and F.M. Fliegel, 
Phys. Rev. B {\bf 24}, 7454 (1981); J.H. Scofield, D.H. Darling and 
W.W. Webb, Phys. Rev. B {\bf 24}, 7450 (1981); J.H. Scofield, 
J.V. Mantese and W.W. Webb, Phys. Rev. B {\bf 32}, 736 (1985); 
D.M. Fleetwood, J.T. Masden and N. Giordano,
Phys. Rev. Lett. {\bf 50}, 450 (1983).

\bibitem{Zhang89} Y.-C. Zhang, Phys. Rev. Lett. {\bf 63}, 470 (1989).

\bibitem{MMM93} S.L. Miller, W.M. Miller and P.J. McWhorter,
J. Appl. Phys. {\bf 73}, 2617 (1993).

\bibitem{McWhorter57} A.L. McWhorter, in {\it Semiconductor Surface Physics},
ed. R.H. Kingston (University of Pennsylvania, Philadelphia, 1957).

\bibitem{MKK90} S. Manna, L.B. Kiss and J. Kertesz, J. Stat. Phys. {\bf 61}, 
923 (1990).

\bibitem{COB92} K. Christensen, Z. Olami and P. Bak, Phys. Rev. Lett. {\bf 68},
2417 (1992).

\bibitem{VZ97} A. Vespignani and S. Zapperi, Phys. Rev. Lett. {\bf 78},
4793 (1997); A. Vespignani and S. Zapperi, Phys. Rev. E {\bf 57}, 6345 (1998).

\end{thebibliography}
\end{document}